\documentstyle[twoside,epsf,psfig,times]{article}
\def\beqa{\begin{eqnarray}}
\def\eeqa{\end{eqnarray}}
\def\beq{\begin{equation}}
\def\eeq{\end{equation}}
\catcode`\@=11
\long\def\@makefntext#1{
\protect\noindent \hbox to 3.2pt {\hskip-.9pt  
$^{{\eightrm\@thefnmark}}$\hfil}#1\hfill}               

\def\@makefnmark{\hbox to 0pt{$^{\@thefnmark}$\hss}}    

\def\ps@myheadings{\let\@mkboth\@gobbletwo
\def\@oddhead{\hbox{}
\rightmark\hfil\eightrm\thepage}   
\def\@oddfoot{}\def\@evenhead{\eightrm\thepage\hfil
\leftmark\hbox{}}\def\@evenfoot{}
\def\sectionmark##1{}\def\subsectionmark##1{}}

\oddsidemargin=\evensidemargin
\newcounter{sectionc}\newcounter{subsectionc}\newcounter{subsubsectionc}
\renewcommand{\section}[1] {\vspace{12pt}\addtocounter{sectionc}{1} 
\setcounter{subsectionc}{0}\setcounter{subsubsectionc}{0}\noindent 
        {\tenbf\thesectionc. #1}\par\vspace{5pt}}
\renewcommand{\subsection}[1] {\vspace{12pt}\addtocounter{subsectionc}{1} 
\setcounter{subsubsectionc}{0}\noindent 
{\bf\thesectionc.\thesubsectionc. {\kern1pt \bfit #1}}\par\vspace{5pt}}
\renewcommand{\subsubsection}[1] {\vspace{12pt}\addtocounter{subsubsectionc}{1}
        \noindent{\tenrm\thesectionc.\thesubsectionc.\thesubsubsectionc.
        {\kern1pt \tenit #1}}\par\vspace{5pt}}
\newcommand{\nonumsection}[1] {\vspace{12pt}\noindent{\tenbf #1}
        \par\vspace{5pt}}

\newcounter{appendixc}
\newcounter{subappendixc}[appendixc]
\newcounter{subsubappendixc}[subappendixc]
\renewcommand{\thesubappendixc}{\Alph{appendixc}.\arabic{subappendixc}}
\renewcommand{\thesubsubappendixc}
        {\Alph{appendixc}.\arabic{subappendixc}.\arabic{subsubappendixc}}

\renewcommand{\appendix}[1] {\vspace{12pt}
        \refstepcounter{appendixc}
        \setcounter{figure}{0}
        \setcounter{table}{0}
        \setcounter{lemma}{0}
        \setcounter{theorem}{0}
        \setcounter{corollary}{0}
        \setcounter{definition}{0}
        \setcounter{equation}{0}
        \renewcommand{\thefigure}{\Alph{appendixc}.\arabic{figure}}
        \renewcommand{\thetable}{\Alph{appendixc}.\arabic{table}}
        \renewcommand{\theappendixc}{\Alph{appendixc}}
        \renewcommand{\thelemma}{\Alph{appendixc}.\arabic{lemma}}
        \renewcommand{\thetheorem}{\Alph{appendixc}.\arabic{theorem}}
        \renewcommand{\thedefinition}{\Alph{appendixc}.\arabic{definition}}
        \renewcommand{\thecorollary}{\Alph{appendixc}.\arabic{corollary}}
        \renewcommand{\theequation}{\Alph{appendixc}.\arabic{equation}}
        \noindent{\tenbf Appendix \theappendixc #1}\par\vspace{5pt}}
\newcommand{\subappendix}[1] {\vspace{12pt}
        \refstepcounter{subappendixc}
        \noindent{\bf Appendix \thesubappendixc. {\kern1pt \bfit #1}}
        \par\vspace{5pt}}
\newcommand{\subsubappendix}[1] {\vspace{12pt}
        \refstepcounter{subsubappendixc}
        \noindent{\rm Appendix \thesubsubappendixc. {\kern1pt \tenit #1}}
        \par\vspace{5pt}}

\newcommand{\textlineskip}{\baselineskip=13pt}
\newcommand{\smalllineskip}{\baselineskip=10pt}

\newcommand{\copyrightheading}[1]
        {\vspace*{-2.5cm}\smalllineskip{\flushleft
        {\footnotesize International Journal of Modern Physics D, #1}\\
        {\footnotesize \copyright\kern2pt World Scientific Publishing
         Company}\\
         }}

\newcommand{\publisher}[2]{{\begin{center}\footnotesize\smalllineskip 
        Received #1\\
        Revised #2
        \end{center}
        }}

\def\abstracts#1#2#3{{
        \centering{\begin{minipage}{4.5in}\footnotesize\baselineskip=10pt
        \parindent=0pt #1\par 
        \parindent=15pt #2\par
        \parindent=15pt #3
        \end{minipage}}\par}} 

\renewenvironment{thebibliography}[1]
        {\frenchspacing
         \ninerm\baselineskip=11pt
         \begin{list}{\arabic{enumi}.}
        {\usecounter{enumi}\setlength{\parsep}{0pt}     
         \setlength{\leftmargin 12.7pt}{\rightmargin 0pt}
         \setlength{\itemsep}{0pt} \settowidth
        {\labelwidth}{#1.}\sloppy}}{\end{list}}

\newcounter{itemlistc}
\newcounter{romanlistc}
\newcounter{alphlistc}
\newcounter{arabiclistc}

\newcommand{\fcaption}[1]{
        \refstepcounter{figure}
        \setbox\@tempboxa = \hbox{\footnotesize Fig.~\thefigure. #1}
        \ifdim \wd\@tempboxa > 5in
           {\begin{center}
        \parbox{5in}{\footnotesize\smalllineskip Fig.~\thefigure. #1}
            \end{center}}
        \else
             {\begin{center}
             {\footnotesize Fig.~\thefigure. #1}
              \end{center}}
        \fi}

\newcommand{\tcaption}[1]{
        \refstepcounter{table}
        \setbox\@tempboxa = \hbox{\footnotesize Table~\thetable. #1}
        \ifdim \wd\@tempboxa > 5in
           {\begin{center}
        \parbox{5in}{\footnotesize\smalllineskip Table~\thetable. #1}
            \end{center}}
        \else
             {\begin{center}
             {\footnotesize Table~\thetable. #1}
              \end{center}}
        \fi}

\def\@citex[#1]#2{\if@filesw\immediate\write\@auxout
        {\string\citation{#2}}\fi
\def\@citea{}\@cite{\@for\@citeb:=#2\do
        {\@citea\def\@citea{,}\@ifundefined
        {b@\@citeb}{{\bf ?}\@warning
        {Citation `\@citeb' on page \thepage \space undefined}}
        {\csname b@\@citeb\endcsname}}}{#1}}

\newif\if@cghi
\def\cite{\@cghitrue\@ifnextchar [{\@tempswatrue
        \@citex}{\@tempswafalse\@citex[]}}
\def\citelow{\@cghifalse\@ifnextchar [{\@tempswatrue
        \@citex}{\@tempswafalse\@citex[]}}
\def\@cite#1#2{{$\null^{#1}$\if@tempswa\typeout
        {IJCGA warning: optional citation argument 
        ignored: `#2'} \fi}}

\def\pmb#1{\setbox0=\hbox{#1}
        \kern-.025em\copy0\kern-\wd0
        \kern.05em\copy0\kern-\wd0
        \kern-.025em\raise.0433em\box0}

\def\fnt#1#2{\footnotetext{\kern-.3em
        {$^{\mbox{\scriptsize #1}}$}{#2}}}

\def\fpage#1{\begingroup
\voffset=.3in
\thispagestyle{empty}\begin{table}[b]\centerline{\footnotesize #1}
        \end{table}\endgroup}

\def\runninghead#1#2{\pagestyle{myheadings}
\markboth{{\protect\footnotesize\it{\quad #1}}\hfill}
{\hfill{\protect\footnotesize\it{#2\quad}}}}
\font\tenrm=cmr10
\font\tenit=cmti10 
\font\tenbf=cmbx10
\font\bfit=cmbxti10 at 10pt
\font\ninerm=cmr9

\font\eightrm=cmr8

\def\qed{\hbox{${\vcenter{\vbox{                  
   \hrule height 0.4pt\hbox{\vrule width 0.4pt height 6pt
   \kern5pt\vrule width 0.4pt}\hrule height 0.4pt}}}$}}

\begin{document}
\setlength{\textheight}{7.7truein}    

\runninghead{Plane-Symmetric Inhomogeneous Bulk Viscous Cosmological Models with Variable $\Lambda$} 
{A. Pradhan and H. R. Pandey}

\normalsize\textlineskip
\thispagestyle{empty}
\setcounter{page}{1}

\copyrightheading{}             {Vol.~0, No.~0 (2002) 000--000}

\vspace*{0.88truein}

\fpage{1}

\centerline{\bf PLANE-SYMMETRIC INHOMOGENEOUS BULK VISCOUS}
\vspace*{0.035truein}
\centerline{\bf COSMOLOGICAL MODELS WITH VARIABLE $\Lambda$}
\vspace*{0.37truein}
\centerline{\footnotesize ANIRUDH PRADHAN\footnote{E-mail: acpradhan@yahoo.com,
pradhan@iucaa.ernet.in (Corresponding Author)}}
\vspace*{0.015truein}  
\centerline{\footnotesize\it Department of Mathematics, Hindu Post-graduate College,}
\baselineskip=10pt
\centerline{\footnotesize\it Zamania, Ghazipur 232 331, India}
\vspace*{10pt}
\centerline{\footnotesize HARE RAM PANDEY}
\vspace*{0.015truein}
\centerline{\footnotesize\it Department of Mathematics, R. S. K. I. College,}
\baselineskip=10pt
\centerline{\footnotesize\it Dubahar, Ballia 277 405, India}
\vspace*{0.225truein}
\publisher{(received date)}{(revised date)}

\vspace*{0.21truein}
\abstracts{A plane-symmetric non-static cosmological model representing a bulk viscous
fluid distribution has been obtained which is inhomogeneous and anisotropic 
and a particular case of which is gravitationally radiative. Without assuming 
any {\it adhoc} law, we obtain a cosmological constant as a decreasing function 
of time. The physical and geometric features of the models are also discussed.}{}{}



\vspace*{1pt}\textlineskip      
\section{Introduction}
\vspace*{-0.5pt}
\noindent
In recent years cosmological models exhibiting plane symmetry have attracted 
the attention of various authors. At the present state of evolution, the universe 
is spherically symmetric and the matter distribution in it is isotropic and 
homogeneous. But in its early stages of evolution, it could have not had  a 
smoothed out picture. Close to the big bang singularity, neither the assumption 
of spherically symmetry nor of isotropy can be strictly valid. So, we consider 
plane symmetry, which is less restrictive than spherical symmetry and 
provides an avenue to study inhomogeneities. Inhomogeneous cosmological cosmological
models play an important role in understanding some essential features of the universe
such as the formation of galaxies during the early stages of evolution and process
of homogenization. The early attempts at the construction of such models have done by 
Tolman\cite{ref1} and Bondi\cite{ref2} who considered spherically symmetric
models. Inhomogeneous plane-symmetric models were considered by Taub\cite{ref3} 
and later by Tomimura,\cite{ref4} Szekeres,\cite{ref5} Collins and Szafron,\cite{ref6}
Szafron and Collins.\cite{ref7} Recently, Senovilla\cite{ref8} obtained a new class 
of exact solutions of Einstein's equation without big bang singularity, representing 
a cylindrically symmetric, inhomogeneous cosmological model filled with perfect fluid
which is smooth and regular everywhere satisfying energy and causality conditions.
Later, Ruis and Senovilla\cite{ref9} have separated out a fairly large class of
singularity free models through a comprehensive study of general cylindrically
symmetric metric with separable function of $r$ and $t$ as metric coefficients.
Dadhich et al.\cite{ref10} have established a link between the FRW model and the 
singularity free family by deducing the latter through a natural and simple
inhomogenization and anisotropization of the former. Recently, Patel et al.\cite{ref11}
presented a general class of inhomogeneous cosmological models filled with 
non-thermalized perfect fluid by assuming that the background spacetime admits two space-like
commuting killing vectors and has separable metric coefficients. Bali and Tyagi\cite{ref12}
obtained a plane-symmetric inhomogeneous cosmological models of perfect fluid distribution
with electromagnetic field. Recently, Pradhan and Yadav\cite{ref13} have investigated
a plane-symmetric inhomogeneous viscous fluid cosmological models with electromagnetic
field.     
\newline
\par
Models with a relic cosmological constant $\Lambda$ have received considerable 
attention recently among researchers for various reasons 
(see Refs.{\cite{ref14}}$^-${\cite{ref18}} and references therein). Some of the 
recent discussions on the cosmological constant ``problem'' and on cosmology 
with a time-varying cosmological constant by Ratra and Peebles,\cite{ref19} 
Dolgov\cite{ref20}$^-$\cite{ref22} and Sahni and Starobinsky\cite{ref23}
point out that in the absence of any interaction with matter or radiation, the 
cosmological constant remains a ``constant'', however, in the presence of
interactions with matter or radiation, a solution of Einstein equations and the 
assumed equation of covariant conservation of stress-energy with a time-varying 
$\Lambda$ can be found. For these solutions, conservation of energy requires 
decrease in the energy density of the vacuum component to be compensated by a 
corresponding increase in the energy density of matter or radiation. Earlier 
researchers on this topic, are contained in Zeldovich,\cite{ref24} 
Weinberg\cite{ref15} and Carroll, Press and Turner.\cite{ref25} Recent
observations by Perlmutter {\it et al.}\cite{ref26} and Riess {\it et al.} \cite{ref27}
strongly favour a significant and positive $\Lambda$. Their finding arise from 
the study of more than $50$ type Ia supernovae with redshifts in the range
$0.10 \leq z \leq 0.83$ and suggest Friedmann models with negative pressure
matter such as a cosmological constant, domain walls or cosmic strings (Vilenkin,
\cite{ref28} Garnavich {\it et al.}\cite{ref29} Recently, Carmeli and Kuzmenko\cite{ref30}
have shown that the cosmological relativity theory (Behar and Carmeli\cite{ref31})
predicts the value $\Lambda = 1.934\times 10^{-35} s^{-2}$ for the cosmological constant.
This value of $\Lambda$ is in excellent agreement with the measurements recently obtained 
by the High-Z Supernova Team and Supernova Cosmological Project (Garnavich et al;\cite{ref29}
Perlmutter et al.;\cite{ref26} Riess et al.;\cite{ref27} Schmidt et al.\cite{ref32}). 
The main conclusion of these works is that the expansion of the universe is accelerating. 
\newline
\par
Several ans$\ddot{a}$tz have been proposed in which the $\Lambda$ term decays 
with time (see Refs. Gasperini,\cite{ref33,ref34} Berman,\cite{ref35} 
Freese {\it et al.},\cite{ref18} $\ddot{O}$zer and Taha,\cite{ref18} 
Peebles and Ratra,\cite{ref36} Chen and Hu,\cite{ref37} Abdussattar and Viswakarma,\cite{ref38}
Gariel and Le Denmat,\cite{ref39} Pradhan {\it et al.}\cite{ref40}). Of the special interest is the
ansatz $\Lambda \propto S^{-2}$ (where $S$ is the scale factor of the
Robertson-Walker metric) by Chen and Wu,\cite{ref37} which has been 
considered/modified by several authors ( Abdel-Rahaman,\cite{ref41} 
Carvalho {\it et al.},\cite{ref18} Waga,\cite{ref42} Silveira and Waga,\cite{ref43}
Vishwakarma.\cite{ref44})
\newline
\par
In most treatments of cosmology, cosmic fluid is considered as perfect fluid.
However, bulk viscosity is expected to play an important role at certain stages
of expanding universe.\cite{ref45}$^-$\cite{ref47} It has been shown that bulk
viscosity leads to inflationary like solution,\cite{ref48} and acts like a negative
energy field in an expanding universe.\cite{ref49} Furthermore, There are several
processes which are expected to give rise to viscous effects. These are the decoupling
of neutrinos during the radiation era and the decoupling of radiation and matter 
during the recombination era. Bulk viscosity is associated with the GUT phase transition 
and string creation. A number of authors have discussed cosmological solutions with 
bulk viscosity in various context.\cite{ref50}$^-$\cite{ref53}
\newline
\par
Roy and Narain\cite{ref54}  have obtained a plane-symmetric non-static cosmological
models in presence of a perfect fluid. In this paper, we will investigate a plane-symmetric 
inhomogeneous bulk viscous fluid cosmological models in the presence of a variable cosmological 
constant varying with time. In our previous paper\cite{ref13} we have obtained a 
non-degenerate Petrov type-II solution. In the present paper we have derived a 
cosmological model, which has in general, as Petrov type-I solution and as a sub case 
it represents a gravitationally radiating Petrov type-II solution. The physical and 
geometric behaviour of the models will be discussed.
\newline
\par

\section{The Metric and Field Equations}
\noindent
We take the plane-symmetric spacetime considered by Roy and Narain\cite{ref54} 
\begin{equation}
\label{eq1} 
 ds^2  = dt^2 - dx^2 - B^2 dy^2 + C^2dz^2, 
\end{equation}
where the metric potentials $B$ and $C$ are functions of $x$ and $t$. The energy 
momentum tensor in the presence of bulk stress has the form 
\begin{equation}
\label{eq2} 
T_{ij} = (\rho + \bar{p}) v_{i} v_{j} - \bar{p} g_{ij}
\end{equation}
and
\begin{equation}
\label{eq3} 
\bar{p} = p - \xi v^{i}_{;i} 
\end{equation}
Here $\rho$, $p$, $\bar{p}$, and $\xi$ are the energy density, isotropic pressure,
effective pressure, bulk viscous coefficient respectively and $v_i$ is the flow
vector satisfying the relation
\begin{equation}
\label{eq4} 
g^{ij} v_{i} v_{j} = 1
\end{equation}
The Einstein's field equations are 
\begin{equation}
\label{eq5} 
R_{ij} - \frac{1}{2}R g_{ij} + \Lambda g_{ij} = - 8\pi T_{ij}
\end{equation}
where $\Lambda$ is the cosmological constant. Eqs. (\ref{eq2}) and (\ref{eq4}) for
the metric (1) lead to
\begin{equation}
\label{eq6} 
v_{2} = v_{3} = 0
\end{equation}
From Eq. (\ref{eq4}), we have
\begin{equation}
\label{eq7} 
v^{2}_{4} - v^{2}_{1} = 1
\end{equation}
The field Eqs. (\ref{eq5}) for the line element (1) lead to
\begin{equation}
\label{eq8} 
-8\pi[(\rho + \bar{p})v^{2}_{1} + \bar{p}] = \frac{B_{44}}{B} + \frac{C_{44}}{C} 
- \frac{B_{1}C_{1}-B_{4}C_{4}}{BC}
\end{equation}
\begin{equation}
\label{eq9} 
-8\pi\bar{p} = \frac{C_{44} - C_{11}}{C} - \Lambda
\end{equation}
\begin{equation}
\label{eq10} 
-8\pi\bar{p} = \frac{B_{44} - B_{11}}{B} - \Lambda
\end{equation}
\begin{equation}
\label{eq11} 
-8\pi[(\rho + \bar{p})v^{2}_{4} - \bar{p}] = \frac{B_{11}}{B} + \frac{C_{11}}{C} 
- \frac{B_{1}C_{1}-B_{4}C_{4}}{BC} + \Lambda
\end{equation}
\begin{equation}
\label{eq12} 
-8\pi[(\rho + \bar{p})v_{1}v_{4}] = \frac{B_{14}}{B} + \frac{C_{14}}{C} 
\end{equation}
The suffixes $1$ and $4$ at the symbols $B$ and $C$ denote partial differential with 
respect to $x$ and $t$ respectively.

\section{Solutions of the Field Equations}
From Eqs. (\ref{eq9}) and (\ref{eq10}), we have
\begin{equation}
\label{eq13} 
\frac{B_{uv}}{B} = \frac{C_{uv}}{C}
\end{equation}
where
\begin{equation}
\label{eq14} 
u = \frac{1}{2}(x + t), ~ ~ v =  \frac{1}{2}(x - t)
\end{equation}
From Eqs. (\ref{eq8})-(\ref{eq13}), we have
\[
\left[\frac{B_{uu} + B_{vv}}{4B} + \frac{C_{uu} + C_{vv}}{4C}\right]^{2} 
- \left[\frac{B_{u}C_{v} + B_{v}C_{u}}{2BC}\right]^{2} \]
\begin{equation}
\label{eq15} 
= \left[\frac{B_{uu} - B_{vv}}{4B} + \frac{C_{uu} - C_{vv}}{4C}\right]^{2}
\end{equation}
Let us assume
\begin{equation}
\label{eq16} 
B = \alpha(u).\beta(v)
\end{equation}
and
\begin{equation}
\label{eq17} 
C = f(u).g(v)
\end{equation}
From Eqs. (\ref{eq13}), (\ref{eq16}) and (\ref{eq17}), we have
\begin{equation}
\label{eq18} 
\frac{(\alpha_{u}/ \alpha)}{(f_{u}/ f)} = \frac{(g_{v}/ g)}{(\beta_{v}/ \beta)} 
= k_{1},
\end{equation}
where $k_{1}$ is an arbitrary constant. Integrating Eq. (\ref{eq18}), we get
\begin{equation}
\label{eq19} 
\alpha = Mf^{k_{1}}
\end{equation}
\begin{equation}
\label{eq20} 
g = N \beta^{k_{1}}
\end{equation}
where $M$ and $N$ are constants of integration. From Eqs. (\ref{eq15}), (\ref{eq16}), 
(\ref{eq17}), (\ref{eq19}) and (\ref{eq20}), we obtain
\begin{equation}
\label{eq21} 
\frac{(f_{u}/f)}{(f_{uu}/f_{u})} = -\left[\frac{(k_{1} + 1)^{2} \frac{\beta_{vv}}{\beta} 
+ k_{1}(k^{2}_{1} -1)\frac{\beta^{2}_{v}}{\beta^{2}}} {k_{1}(k^{2}_{1} - 1)
\frac{\beta_{vv}}{\beta} + \{k^{2}_{1}(k^{2}_{1} - 1)^{2} - (k^{2}_{1} + 1)^{2}\} 
\frac{\beta^{2}_{v}}{\beta^{2}}}\right] = k_{2},
\end{equation}
where $k_{2}$ is an arbitrary constant. Integrating Eq. (\ref{eq21}), we get
\begin{equation}
\label{eq22}
f = (au +b)^{\mu}
\end{equation}
and
\begin{equation}
\label{eq23}
\beta = (mv + n )^{\nu}
\end{equation}
where $a$, $b$, $m$ and $n$ are arbitrary constants and 
\begin{equation}
\label{eq24}
\mu = \frac{k_{2}}{k_{2} - 1}, ~ ~ \nu = \frac{k_{1} + 1}{k^{2}_{1} +1} - \frac{k_{2}}{k_{2} - 1}
\end{equation}
By suitable transformation metric (1) reduces to the form
\begin{equation}
\label{eq25}
ds^{2} = dT^{2} - dX^{2} - {U}^{2k_{1}\mu}{V}^{2\nu}dY^{2} - {U}^{2\mu}{V}^{2k_{1}\nu} dZ^{2}
\end{equation}
where
\begin{equation}
\label{eq26}
U = \frac{1}{2}(X + T) ~ ~ and ~ ~ V = \frac{1}{2}(X - T) 
\end{equation}
The effective pressure and the energy density for the model (25) are given by
\begin{equation}
\label{eq27}
8 \bar{p} = - \left[\frac{4k_{1}\mu\nu}{(T^{2} - X^{2})}\right] + \Lambda
\end{equation}
\begin{equation}
\label{eq28}
8\pi\rho = \left[\frac{4\mu\nu(k^{2}_{1} + k_{1} + 1)}{(T^{2} - X^{2})}\right] - \Lambda
\end{equation}
For the simplification of $\xi$, we assume that the fluid obeys an equation of 
state of the form
\begin{equation}
\label{eq29}
p = \gamma \rho,
\end{equation}
where $\gamma (0\leq \gamma\leq1)$ is a constant.\\
Thus, given $\xi(t)$ we can solve the cosmological parameters. In most of the 
investigations involving bulk viscosity is assumed to be a simple power function
of the energy density\cite{ref55}$^-$\cite{ref57}
\begin{equation}
\label{eq30}
\xi(t) = \xi_{0} \rho^{k}
\end{equation}
where $\xi_{0}$ and $k$ are constants. If $k=1$, Eq. (\ref{eq30}) may correspond
to a radiative fluid. However, more realistic models\cite{ref58} are based on $n$ 
lying in the regime $0\leq n\leq\frac{1}{2}$.

\subsection{Solutions for $\xi = \xi_{0}$}
In this case we assume $k = 0$ in Eq. (\ref{eq30}). Eqs. (\ref{eq30}) and (\ref{eq27}) 
become  $\xi = \xi_{0}$ = constant and
\begin{equation}
\label{eq31}
8\pi p = \frac{8\pi\xi_{0}}{(T^{2} - X^{2})} \left[1 + \frac{(k_{1} + 1)^{2}}{(k^{2}_{1} + 1)}\right]
- \frac{4k_{1}\mu \nu}{(T^{2} - X^{2})} + \Lambda
\end{equation}
Using Eq. (\ref{eq29}) and eliminating $\rho(t)$ from Eqs. (\ref{eq28}) and (\ref{eq31})
we obtain
\begin{equation}
\label{eq32}
\Lambda = \frac{2\mu \nu(k_{1} + 1)^{2}}{(T^{2} - X^{2})} -
\frac{4\pi\xi_{0}}{(T^{2} - X^{2})^{\frac{1}{2}}}\left[1 + \frac{(k_{1} + 1)^{2}}{(k^{2}_{1} + 1)}\right]
\end{equation}

\subsection{Solutions for $\xi = \xi_{0}\rho$}
In this case we assume $k = 1$ in Eq. (\ref{eq30}). Hence, Eqs. (\ref{eq30}) and
(\ref{eq27}) become
\begin{equation}
\label{eq33}
\xi = \xi_{0} \rho
\end{equation}
and
\begin{equation}
\label{eq34}
8\pi \left[p - \frac{\xi_{0}\rho}{(T^{2} - X^{2})^{\frac{1}{2}}}\left(1 + \frac{(k_{1} + 1)^{2}}
{(k^{2}_{1} + 1)}\right)\right] = - \frac{4k_{1} \mu \nu}{(T^{2} - X^{2})} + \Lambda 
\end{equation}
Using Eq. (\ref{eq29}) and eliminating $\rho(t)$ from Eqs. (\ref{eq28})
and (\ref{eq34}) we obtain
\[
\left[1 + \gamma - \frac{\xi_{0}}{(T^{2} - X^{2})^{\frac{1}{2}}}\left(1 +  \frac{(k_{1} + 1)^{2}}
{(k^{2}_{1} + 1)}\right)\right]\Lambda = \]
\begin{equation}
\label{eq35}
\frac{4\mu\nu}{(T^{2} - X^{2})}\left[k_{1} + (k^{2}_{1} + k_{1} + 1)\left(\gamma -
\frac{\xi_{0}}{(T^{2} - X^{2})^{\frac{1}{2}}}\{1 + \frac{(k_{1} + 1)^{2}}{(k^{2}_{1} + 1)}
\}\right)\right]
\end{equation}
We have observed from Eqs. (\ref{eq32}) and (\ref{eq35}) that the cosmological constant
is a decreasing function of time and it approaches a small value as time 
progresses (i.e., the present epoch). Thus, with our approach, we obtain a physically
relevant decay law for the cosmological constant unlike other investigators where 
{\it adhoc} laws were used to arrive at a mathematical expression for the decaying 
vacuum energy.

\section {Some Physical and Geometrical Features of the Models}
We shall now give the expressions for kinematical quantities and components of 
conformal curvature tensor. With regard to the kinematical properties of the 
velocity vector $v_{i}$ in the metric (25), a straight forward calculation leads
to the following expressions for the expansion $\theta$ and shear tensor $\sigma_{ij}$
of the fluid. 
\begin{equation}
\label{eq36}
\theta = \frac{1}{(T^{2} - X^{2})^{\frac{1}{2}}}
\left[1 + \frac{(k_{1} + 1)^{2}}{(k^{2}_{1} + 1)}\right]
\end{equation}
\begin{equation}
\label{eq37}
\sigma_{11} = \frac{2k_{1}}{3(k^{2}_{1} + 1)}
\left[\frac{T^{2}}{(T^{2} - X^{2})^{\frac{3}{2}}}\right]
\end{equation}
\begin{equation}
\label{eq38}
\sigma_{22} = \frac{U^{2k_{1}\mu} V^{2\nu}}{(T^{2} - X^{2})^{\frac{1}{2}}}
\left[\frac{2k^{2}_{1} - k_{1} - 1}{3(k^{2}_{1} + 1)} - \mu(k_{1} - 1)\right] 
\end{equation}
\begin{equation}
\label{eq39}
\sigma_{33} = \frac{U^{2\mu} V^{2k_{1}\nu}}{(T^{2} - X^{2})^{\frac{1}{2}}}
\left[\frac{2 - k_{1}(k_{1} + 1)}{3(k^{2}_{1} + 1)} - \mu(k_{1} - 1)\right] 
\end{equation}
\begin{equation}
\label{eq40}
\sigma_{44} = - \left[\frac{(k_{1} - 1)^{2}}{3(k^{2}_{1} + 1)}\right]
\left[\frac{X^{2}}{(T^{2} - X^{2})^{\frac{3}{2}}}\right]
\end{equation}
\begin{equation}
\label{eq41}
\sigma_{14} = \left[\frac{(k_{1} - 1)^{2}}{3(k^{2}_{1} + 1)}\right]
\left[\frac{TX}{(T^{2} - X^{2})^{\frac{3}{2}}}\right]
\end{equation}
where all other components, the rotational tensor and the acceleration vanish.
Hence the models are expanding, non-rotating, shearing and geodetic in general.
The non-vanishing components of the flow vector are
\begin{equation}
\label{eq42}
v_{1} = -\frac{X}{(T^{2} - X^{2})^{\frac{1}{2}}}
\end{equation}
\begin{equation}
\label{eq43}
v_{4} = \frac{T}{(T^{2} - X^{2})^{\frac{1}{2}}}
\end{equation}
It is observed that the region of the spacetime in which this is valid, 
is $T^{2} - X^{2} > 0$. The reality conditions $p>0$ and $\rho>0$ imply that
\begin{equation}
\label{eq44} 
\left[\frac{4k_{1}\mu\nu}{(T^{2} - X^{2})}\right] < \Lambda < 
\left[\frac{2\mu\nu (k_{1} + 1)^{2}}{(T^{2} - X^{2})}\right]
\end{equation}
The non-vanishing components of the conformal curvature tensor are
\[
C_{(1212)} = \frac{1}{8}\left[\frac{\mu(k_{1} - 1)\{\mu(k_{1} + 1) - 1\}}{U^{2}}
-\frac{\nu(k_{1} - 1)\{\nu(k_{1} + 1) - 1\}}{V^{2}}\right] \]
\begin{equation}
\label{eq45} 
- \frac{1}{12}\left[\frac{\mu\nu(k_{1} - 1 )^{2}}{UV}\right]
\end{equation}
\[
C_{(1313)} = \frac{1}{8}\left[\frac{\nu(k_{1} - 1)\{\nu(k_{1} + 1) - 1\}}{V^{2}}
-\frac{\mu(k_{1} - 1)\{\mu(k_{1} + 1) - 1\}}{U^{2}}\right] \]
\begin{equation}
\label{eq46} 
- \frac{1}{12}\left[\frac{\mu\nu(k_{1} - 1 )^{2}}{UV}\right]
\end{equation}
\begin{equation}
\label{eq47} 
C_{(2323)} = \frac{1}{6}\left[\frac{\mu\nu(k_{1} - 1 )^{2}}{UV}\right]
\end{equation}
\[
C_{(2424)} = \frac{1}{8}\left[\frac{\mu(k_{1} - 1)\{\mu(k_{1} + 1) - 1\}}{U^{2}}
-\frac{\nu(k_{1} - 1)\{\nu(k_{1} + 1) - 1\}}{V^{2}}\right] \]
\begin{equation}
\label{eq48} 
+ \frac{1}{12}\left[\frac{\mu\nu(k_{1} - 1 )^{2}}{UV}\right]
\end{equation}
\[
C_{(3434)} = \frac{1}{8}\left[\frac{\nu(k_{1} - 1)\{\nu(k_{1} + 1) - 1\}}{V^{2}}
-\frac{\mu(k_{1} - 1)\{\mu(k_{1} + 1) - 1\}}{U^{2}}\right] \]
\begin{equation}
\label{eq49} 
+ \frac{1}{12}\left[\frac{\mu\nu(k_{1} - 1 )^{2}}{UV}\right]
\end{equation}
\begin{equation}
\label{eq50} 
C_{(1414)} = - \frac{1}{6}\left[\frac{\mu\nu(k_{1} - 1 )^{2}}{UV}\right]
\end{equation}
\begin{equation}
\label{eq51} 
C_{(1224)} = - \frac{1}{8}\left[\frac{\mu(k_{1} - 1)\{\mu(k_{1} + 1) - 1\}}{U^{2}}
-\frac{\nu(k_{1} - 1)\{\nu(k_{1} - 1) - 1\}}{V^{2}}\right] 
\end{equation}
\begin{equation}
\label{eq52} 
C_{(1334)} = \frac{1}{8}\left[\frac{\mu(k_{1} - 1)\{\mu(k_{1} + 1) - 1\}}{U^{2}}
+\frac{\nu(k_{1} - 1)\{\nu(k_{1} + 1) - 1\}}{V^{2}}\right] 
\end{equation}
From Eqs. (\ref{eq45}) - (\ref{eq52}) it is observed that if $k_{1} = 1$ 
the spacetime is conformally flat. The spacetime is, in general, of 
Petrov-type I. However if $k_{1} \ne 1$ and either $\mu = \frac{1}{k_{1} + 1}$ 
or $\nu = \frac{1}{k_{1} + 1}$ then the spacetime will be of Petrov-type II. 
For this case the reality conditions $p > 0$ and $\rho > 0$ imply that
\begin{equation}
\label{eq53}
\left[\frac{8 k^{2}_{1}}{(k_{1} + 1)^{2}(k^{2}_{1} + 1)(T^{2} - X^{2})}\right]
< \Lambda < \left[\frac{4 k_{1}}{(k^{2}_{1} + 1)(T^{2} - X^{2})}\right]
\end{equation}
For $\mu = \frac{1}{k_{1} + 1}$, we have $\nu = \frac{2k_{1}}{(k_{1} + 1)(k^{2}_{1} + 1)}$
and in this case the geometry of the spacetime (25) takes the form 
\begin{equation}
\label{eq54}
ds^{2} = dT^{2} - dX^{2} - {U}^{\frac{2k_{1}}{(k_{1} + 1)}} {V}^{\frac{2k_{1}}{(k_{1} + 1)
(k^{2}_{1} + 1)}} dY^{2} - {U}^{\frac{2}{(k_{1} + 1)}} {V}^{\frac{4k^{2}_{1}}{(k_{1} + 1)
(k^{2}_{1} + 1)}} dZ^{2} 
\end{equation}
The physical components $C_{hijk}$ take the form
\begin{equation}
\label{eq55}
\mathbf{C_{AB}}=
\left( \begin{array}{cccccc}
-2\xi & 0 & 0 & 0 & 0 & 0 \\
0 & \xi-\eta & 0 & 0 & 0 & 0 \\
0 & 0 & \xi + \eta & 0 & \eta & 0 \\
0 & 0 & 0 & 2\xi & 0 & 0 \\
0 & 0 & \eta & 0 & -\xi + \eta & 0 \\
0 & \eta & 0 & 0 & 0 & -\xi-\eta \\
\end{array} \right)
\end{equation}
where
\begin{equation}
\label{eq56}
\xi = - \frac{1}{3}\left[\frac{(k_{1} - 1)^{2}}{(k_{1} + 1)^{2}(k^{2}_{1} + 1)}\right]
\frac{1}{UV}
\end{equation}
and
\begin{equation}
\label{eq57}
\eta = \frac{1}{4}\left[\frac{k^{3}_{1}(k_{1} - 1)}{(k_{1} + 1)(k^{2}_{1} + 1)^{2}}\right]
\frac{1}{V^{2}}
\end{equation}
From Eq. (\ref{eq55}) we conclude that the model exhibits gravitational radiation.
 For large value of $X$ it gives a type-two null spacetime representing an
outgoing radiation field, although it will not satisfy the reality conditions
at $X = \infty$. The radiative term in $C_{AB}$ is the quantity\\
$\eta = \frac{1}{4}\left[\frac{k^{3}_{1}(k_{1} - 1)}{(k_{1} + 1)(k^{2}_{1} + 1)^{2}}\right]
\frac{1}{V^{2}}.$\\

\section{Discussion and Concluding Remarks}
\noindent We have obtained a new class of inhomogeneous plane-symmetric cosmological
models with a bulk viscous fluid as the source of matter. Generally, the models
are expanding, shearing, non-rotating and Petrov-type I non-degenerate
in which the flow vector is geodetic. Under certain conditions the models will 
be of Petrov-type II. It is well known fact that the free gravitational field 
affects the flow of the fluid by inducing the shear in the flow line.\cite{ref63}
It is therefore interesting to study the cosmological models with given Petrov-types.
The cosmological constant in all models given in Sec. 3.1 and 3.2 are decreasing 
function of time. The cosmological consequences of a decaying cosmological term has 
been discussed lucidly\cite{ref35},\cite{ref59}$^-$\cite{ref62} with an {\it adhoc} 
assumption of a decay law. However, in our models, we recover such a law without 
assuming any {\it a priori} law for $\Lambda$. Thus our models are more general than 
those studied earlier.\\ 

\nonumsection{Acknowledgements}
\noindent One of the authors (A. Pradhan) wishes to thank the Inter-University 
Centre for Astronomy and Astrophysics, Pune, India for providing facility 
under Associateship Programme where this work was done. We would like to thank 
R. G. Vishwakarma for many helpful discussions.
\newline
\newline
\nonumsection{References}


\begin{thebibliography}{000}
\bibitem {ref1} 
R. C. Tolman, {\it Proc. Nat. Acad. Sci.} {\bf 20}, 169 (1934).
\bibitem {ref2}  
H. Bondi, {\it Mon. Not. R. Astro. Soc.} {\bf 107}, 410 (1947). 
\bibitem {ref3} 
A. H. Taub, {\it Ann. Math.} {\bf 53}, 472 (1951); {\it Phys. Rev.} {\bf 103}, 454 (1956).
\bibitem {ref4}  
N. Tomimura, {\t II Nuovo Cimento} {\bf B44}, 372 (1978)
\bibitem {ref5} 
P. Szekeres, {\it Commun. Math. Phys.} {\bf 41}, 55 (1975). 
\bibitem {ref6}  
C. B. Collins and D. A. Szafron, {\it J. Math. Phy.} {\bf 20}, 2347 (1979a);\\
{\it J. Math. Phy.} {\bf 20}, 2362 (1979b).
\bibitem {ref7}
D. A. Szafron and C. B. Collins, {\it J. Math. Phy.} {\bf 20}, 2354 (1979)
\bibitem {ref8}
J. M. M. Senovilla, {\it Phy. Rev. Lett.} {\bf 64}, 2219 (1990). 
\bibitem {ref9}
E. Ruiz and J. M. M. Senovilla, {\it Phy. Rev.} {\bf D45}, 1995 (1990). 
\bibitem {ref10}
N. Dadhich, R. Tikekar and L. K. Patel, {\it Curr. Sci.} {\bf 65}, 694 (1993).
\bibitem {ref11}
L. K. Patel, R. Tikekar and N. Dadhich, {\it Pramana-J. Phys.} {\bf 49}, 213 (1993).  
\bibitem {ref12}
R. Bali and A. Tyagi, {\it Astrophys. Space Sc.} {\bf 173}, 233 (1990).
\bibitem {ref13}
A. Pradhan, V. K. Yadav and N. N. Saste, {\it Int. J. Mod. Phys.} {\bf D11}, 857 (2002).
\bibitem {ref14}
S. Weinberg, {\it Rev. Mod. Phys.} {\bf 61}, 1 (1989).
\bibitem {ref15}
S. Weinberg, {\it Gravitation and Cosmology} (Wiley, New York, 1972).
\bibitem {ref16}
J. A. Frieman and I. Waga, {\it Phys. Rev.} D {\bf 57}, 4642 (1998).
\bibitem {ref17}
R. Carlberg, {\it et al}. {\it Astrophys. J.} {\bf 462}, 32 (1996).
\bibitem {ref18}
M. $\ddot{O}$zer and M. O. Taha, {\it Nucl. Phys.} B {\bf 287}, 776 (1987);\\
K. Freese, F. C. Adams, J. A. Frieman and E. Motta, {\it ibid.} B {\bf 287}, 
1797 (1987);\\
J. C. Carvalho, J. A. S. Lima and I. Waga, {\it Phys. Rev.} D {\bf 46}, 2404 
(1992);\\
V. Silviera and I. Waga., {\it ibid.} D {\bf 50}, 4890 (1994).
\bibitem {ref19}
B. Ratra and P. J. E. Peebles, {\it Phys. Rev.} D {\bf 37}, 3406 (1988).
\bibitem {ref20}
A. D. Dolgov, in {\it The Very Early Universe}, eds. G. W. Gibbons, S. W. 
Hawking and S. T. C. Siklos (Cambridge Univerity Press, 1983).
\bibitem {ref21}
A. D. Dolgov, M. V. Sazhin and Ya. B. Zeldovich, {\it Basics of Modern 
Cosmology} (Editions Frontiers, 1990).
\bibitem {ref22}
A. D. Dolgov, {\it Phys. Rev.} D {\bf 55}, 5881 (1997).
\bibitem {ref23}
V. Sahni and A. Starobinsky, {\it Int. J. Mod. Phys.} D {\bf 9}, 373 (2000). 
(gr-qc/9904398) (2000).
\bibitem {ref24}
Ya. B. Zeldovich, {\it Sov. Phys.-Uspekhi} {\bf 11}, 381 (1968).
\bibitem {ref25}
S. M. Carroll, W. H. Press and E. L. Turner, {\it Ann. Rev. Astron. Astrophys.} 
{\bf 30}, 499 (1992).
\bibitem {ref26}
S. Perlmutter {\it et al.}, {\it Astrophys. J.} {\bf 483}, 565 (1997), Supernova Cosmology 
Project Collaboration (astro-ph/9608192); 
{\it Nature} {\bf 391}, 51 (1998), Supernova Cosmology 
Project Collaboration (astro-ph/9712212); 
{\it Astrophys. J.} {\bf 517}, 565 (1999), 
Project Collaboration (astro-ph/9608192).
\bibitem {ref27}
A. G. Riess {\it et al.}, {\it Astron. J.} {\bf 116}, 1009 (1998); Hi-Z Supernova Team Collaboration
(astro-ph/9805201).
\bibitem {ref28}
A. Vilenkin, {\it Phys. Rep.} {\bf 121}, 265 (1985).
\bibitem {ref29}
P. M. Garnavich {\it et al.}, {\it Astrophys. J.} {\bf 493}, L53 (1998a),
Hi-Z Supernova Team Collaboration (astro-ph/9710123); 
{\it Astrophys. J.} {\bf 509}, 74 (1998b);
Hi-Z Supernova Team Collaboration (astro-ph/9806396).
\bibitem {ref30}
M. Carmeli and T. Kuzmenko, {\it Int. J. Theor. Phys.} {\bf 41}, 131 (2002).
\bibitem {ref31}
S. Behar and M. Carmeli, {\it Int. J. Theor. Phys.} {\bf 39}, 1375 (2002); astro-ph/0008352.
\bibitem {ref32}
B. P. Schmidt {\it et al.}, {\it Astrophys. J.} {\bf 507}, 46 (1998),
Hi-Z Supernova Team Collaboration (astro-ph/9805200). 
\bibitem {ref33}
M. Gasperini, {\it Phys. Lett. } B {\bf 194}, 347 (1987).
\bibitem {ref34}
M. Gasperini, {\it Class. Quantum Grav.} {\bf 5}, 521 (1988).
\bibitem {ref35}
M. S. Berman, {\it Int. J. Theor. Phys.} {\bf 29}, 567 (1990); 
{\it Int. J. Theor. Phys.} {\bf 29}, 1419 (1990); {\it Phys. Rev.} {\bf D43}, 75 (1991);\\
M. S. Berman and M. M. Som, {\it Int. J. Theor. Phys.} {\bf 29}, 1411 (1990);\\
M. S. Berman, M. M. Som and F. M. Gomide, {\it Gen. Rel. Gravit.} {\bf 21}, 287 (1989);\\
M. S. Berman and F. M. Gomide, {\it Gen. Rel. Gravit.} {\bf 22}, 625 (1990).  
\bibitem {ref36}
P. J. E. Peebles and B. Ratra, {\it Astrophys. J.} {\bf 325}, L17 (1988).
\bibitem {ref37}
W. Chen and Y. S. Wu, {\it Phys. Rev.}  {\bf D41}, 695 (1990).
\bibitem {ref38}
Abdussattar and R. G. Vishwakarma, {\it Pramana J. Phys.} {\bf 47}, 41 (1996).
\bibitem {ref39}
J. Gariel and G. Le Denmat, {\it Class. Quant. Grav.} {\bf 16}, 149 (1999).
\bibitem {ref40}
A. Pradhan and A. Kumar, {\it Int. J. Mod. Phys.}  {\bf D10}, 291 (2001);\\
A. Pradhan and V. K. Yadav, {\it Int J. Mod Phys.} {\bf D11}, 839 (2002).
\bibitem {ref41}
A.-M. M. Abdel-Rahaman, {\it Gen. Rel. Grav.} {\bf 22}, 655 (1990); {\it Phys. Rev. }  
{\bf D45}, 3492 (1992). 
\bibitem {ref42}
I. Waga, {\it Astrophys. J.} {\bf 414}, 436 (1993).
\bibitem {ref43}
V. Silveira and I. Waga, {\it Phys. Rev.}  {\bf D50}, 4890 (1994).
\bibitem {ref44}
R. G. Vishwakarma, {\it Class. Quantum Grav.} {\bf 17}, 3833 (2000).
\bibitem {ref45}
C. W. Misner, {\it astrophys. J.} {\bf 151}, 431 (1968).
\bibitem {ref46}
G. F. R. Ellis, In {\it General Relativity and Cosmology}, Enrico Fermi Course, 
R. K. Sachs. ed. (Academic, New York, 1979).
\bibitem {ref47}
B. L. Hu, In {\it Advance in Astrophysics}, eds. L. J. Fung and R. Ruffini,
(World Scientific, Singapore, 1983).
\bibitem {ref48}
T. Padmanabhan and S. M. Chitre, {\it Phys. Lett.} {\bf A120}, 433 (1987).
\bibitem {ref49}
V. B. Johri and R. Sudarshan, {\it Proc. Int. Conf. on Mathematical Modelling in 
Science and Technology}, L. S. Srinath et al., eds (World Scientific, Singapore, 1989).
\bibitem {ref50}
\O. Gr\o n, {\it Astrophys. Space Sci.} {\bf 173}, 191 (1990).
\bibitem {ref51}
A. Pradhan, V. K. Yadav and I. Chakrabarty, {\it Int J. Mod. Phys.} {\bf D10}, 339 (2001);\\
I. Chakrabarty, A. Pradhan and N. N. Saste, {\it Int J. Mod. Phys.} {\bf D10}, 741 (2001); \\
A. Pradhan and I. Aotemshi, {\it Int J. Mod. Phys.} {\bf D11}, (2002), in press. 
\bibitem {ref52}
L. P. Chimento, A. S. Jakubi and D. Pavon, {\it Class. Quantum Grav.} {\bf 16}, 1625 (1999).
\bibitem {ref53}
G. P. Singh, S. G. Ghosh and A. Beesham, {\it Aust. J. Phys.} {\bf 50}, 903 (1997).
\bibitem {ref54}
S. R. Roy and S. Narain, {\it Indian J. pure appl. Math.} {\bf 12}, 284 (1981).
\bibitem {ref55}
D. Pavon, J. Bafaluy and D. Jou, {\it Class. Quantum Gravit.} {\bf 8}, 357 (1991);
{\it Proc. Hanno Rund Conf. on Relativity and Thermodynamics}, Ed. S. D. Maharaj, (University of Natal,
Durban, 1996), p. 21.
\bibitem {ref56}
R. Maartens, {\it Class. Quantum Gravit.} {\bf 12}, 1455 (1995).
\bibitem {ref57}
W. Zimdahl, {\it Phys Rev.} {\bf D53}, 5483 (1996).
\bibitem {ref58}
N. O. Santos, R. S. Dias and A. Banerjee, {\it J. Math. Phys.} {\bf 26}, 878 (1985).
\bibitem {ref59}
A. Beesham, {\it Int. J. Theor. Phys.} {\bf 25}, 1295 (1986).
\bibitem {ref60}
T. Singh, A. Beesham and W. S. Mbokazi, {\it Gen. Rel. Gravit.} {\bf 30}, 537 (1998). 
\bibitem {ref61}
M. $\ddot{O}$zer and M. O. Taha, {\it Phys Lett.} {\bf B171}, 363 (1986). 
\bibitem {ref62}
M. V. John and K. B. Joseph, {\it Phys Rev.} {\bf D61}, 087304 (2000). 
\bibitem {ref63}
G. F. R. Ellis, {\it General Relativity and Gravitation}, R. K. Sachs, ed. 
(Academic Press, New York, 1971), p. 129.
\end{thebibliography}
\end{document}